\documentclass[conference]{IEEEtran}
\IEEEoverridecommandlockouts
\usepackage{cite}
\usepackage{amsmath,amssymb,amsfonts}
\usepackage{algorithmic}
\usepackage{graphicx}
\usepackage{textcomp}
\usepackage{xcolor}
\usepackage{multirow}
\usepackage[flushleft]{threeparttable}
\usepackage{float}
\usepackage{subfig}
\usepackage[none]{hyphenat}
\usepackage{physics}
\usepackage{siunitx}
\usepackage{makecell}
\usepackage{dblfloatfix}
\usepackage{url}
\usepackage{soul}
\usepackage[utf8]{inputenc}  

\definecolor{deepgreen}{RGB}{34,139,34} 
\usepackage[a4paper, total={184mm,239mm}]{geometry}
\def\BibTeX{{\rm B\kern-.05em{\sc i\kern-.025em b}\kern-.08em
    T\kern-.1667em\lower.7ex\hbox{E}\kern-.125emX}}
\begin{document}
\title{There's Waldo: PCB Tamper Forensic Analysis using  Explainable AI on Impedance Signatures

\author{
  Maryam Saadat Safa\textsuperscript{*}\thanks{*These authors contributed equally to this work. 
  
  This is the author version of the paper accepted for presentation at 2025 IEEE International Symposium On Electromagnetic Compatibility, Signal \& Power Integrity. }, 
  Seyedmohammad Nouraniboosjin\textsuperscript{*}, 
  Fatemeh Ganji, 
  and Shahin Tajik\\
  Worcester Polytechnic Institute, Worcester, MA, USA\\
  \{msafa, snouraniboosjin, fganji, stajik\}@wpi.edu
}
}
\maketitle

\begin{abstract}
The security of printed circuit boards (PCBs) has become increasingly vital as supply chain vulnerabilities, including tampering, present significant risks to electronic systems.
While detecting tampering on a PCB is the first step for
verification, forensics is also needed to identify
the modified component.
One non-invasive and reliable PCB tamper detection technique with global coverage is the impedance characterization of PCB's power delivery network (PDN).
However, it is an open question whether one can use the two-dimensional impedance signatures for forensics purposes.
In this work, we introduce a novel PCB forensics approach, using explainable AI (XAI) on impedance signatures. 
Through extensive experiments, we replicate various PCB tamper events, generating a dataset used to develop an XAI algorithm capable of not only detecting tampering but also explaining why the algorithm makes a decision about whether a tamper event has happened.  
At the core of our XAI algorithm is a random forest classifier with an accuracy of 96.7\%, sufficient to explain the algorithm's decisions.   
To understand the behavior of the classifier 
In the decision-making process, we utilized the SHAP values as an XAI tool to determine which frequency component influences the classifier's decision for a particular class the most. 
This approach enhances detection capabilities as well as advancing the verifier's ability to reverse-engineer and analyze two-dimensional impedance signatures for forensics.
\end{abstract}


\section{Introduction}
Modern PCBs, designed with high-density interconnects, tight component spacing, and multiple routing layers, are increasingly in demand; however, the ability of manufacturers to meet these requirements has diminished, leading to a reliance on a globalized supply chain.
This shift has led to an increased reliance on potentially compromised third-party components, contract manufacturers, and electronic design automation (EDA) tools, raising concerns about the integrity and security of the final products.
This globalization intensifies the challenge of maintaining security and integrity, as traditional detection and protection mechanisms are bypassed, making PCBs susceptible to tampering, hardware Trojans, and counterfeit components.

A reliable, cost-effective method for detecting such tampering attacks is impedance characterization of the PCB's power delivery network~\cite{mosavirik2022scatterverif,werner2022detection,zhu2023pdnpulse,safa2023counterfeit,mosavirik2023impedanceverif}.
Since any tampering with the PCB or IC package results in changes to the equivalent impedance of the power delivery network (PDN), characterizing it over a range of frequency bands reveals if the physical integrity of the system has been violated.
To characterize the PDN’s impedance over frequency, S-(scattering) and Z-(Impedance) parameters
are deployed.
While detecting tampering on a system is an essential initial step in verification, it is equally important to identify the modified component and its location within the system. 
Currently, S-/Z-parameters are primarily used for detection. However, to determine the root cause of deviations in these parameters, additional inspection methods, such as visual inspection, are necessary.
These supplementary inspections can be both costly and time-consuming. 
Hence, it would be ideal if the verifier could reverse-engineer and disassemble the two-dimensional S-/Z-parameters for forensic purposes.
However, modern multi-layered PCBs present significant challenges due to the complex electromagnetic interactions among components (e.g., traces, vias, and planes), leading to coupled equations that are difficult to resolve manually. 
Hence, we ask the following research question: \emph{To what degree is it possible to reverse engineer
the S-/Z-parameter signatures to determine the root cause and location of tamper events?}


\noindent{\bfseries Our Contribution:}
The paper aims to take a step towards enabling the verifier to reverse-engineer and disassemble the two-dimensional (amplitude and frequency) S-/Z-parameters for forensic analysis and enabling a more profound understanding of the impact of tampering.
Hence, we present a novel approach for PCB forensics by applying explainable AI (XAI) to impedance signatures. Through extensive experiments, we simulate various PCB tamper events, creating a dataset to develop an XAI algorithm that not only detects tampering but also explains the reasoning behind its decisions. The core of our XAI algorithm is a random forest classifier, achieving 96.7\% accuracy, sufficient for providing explainability.
To understand the classifier's decision-making process, we used SHAP values, enabling us to identify which frequency components have the most influence on the classifier's decisions for specific classes. This method improves tamper detection while enhancing the ability to reverse-engineer and analyze two-dimensional impedance signatures for forensic applications. 

\section{Background}\label{background}
\subsection{Power Delivery Network}
The Power Delivery Network (PDN) plays a critical role in providing a stable supply voltage to various modules on a PCB. 
The PDN consists of various electronic components and interconnects, including voltage regulator modules, decoupling capacitors, and power rails on the chip.
While at lower frequencies, the PDN impedance is dominated by the voltage regulator, at higher frequencies, the off-chip/on-chip components contribute most to it.
The low impedance of the PDN makes it highly sensitive to even minor modifications~\cite{zhu2023pdnpulse}.

\subsection{Scattering Parameter}
\begin{figure}[t!]
     	\centering \noindent
     	\includegraphics[width=8cm]{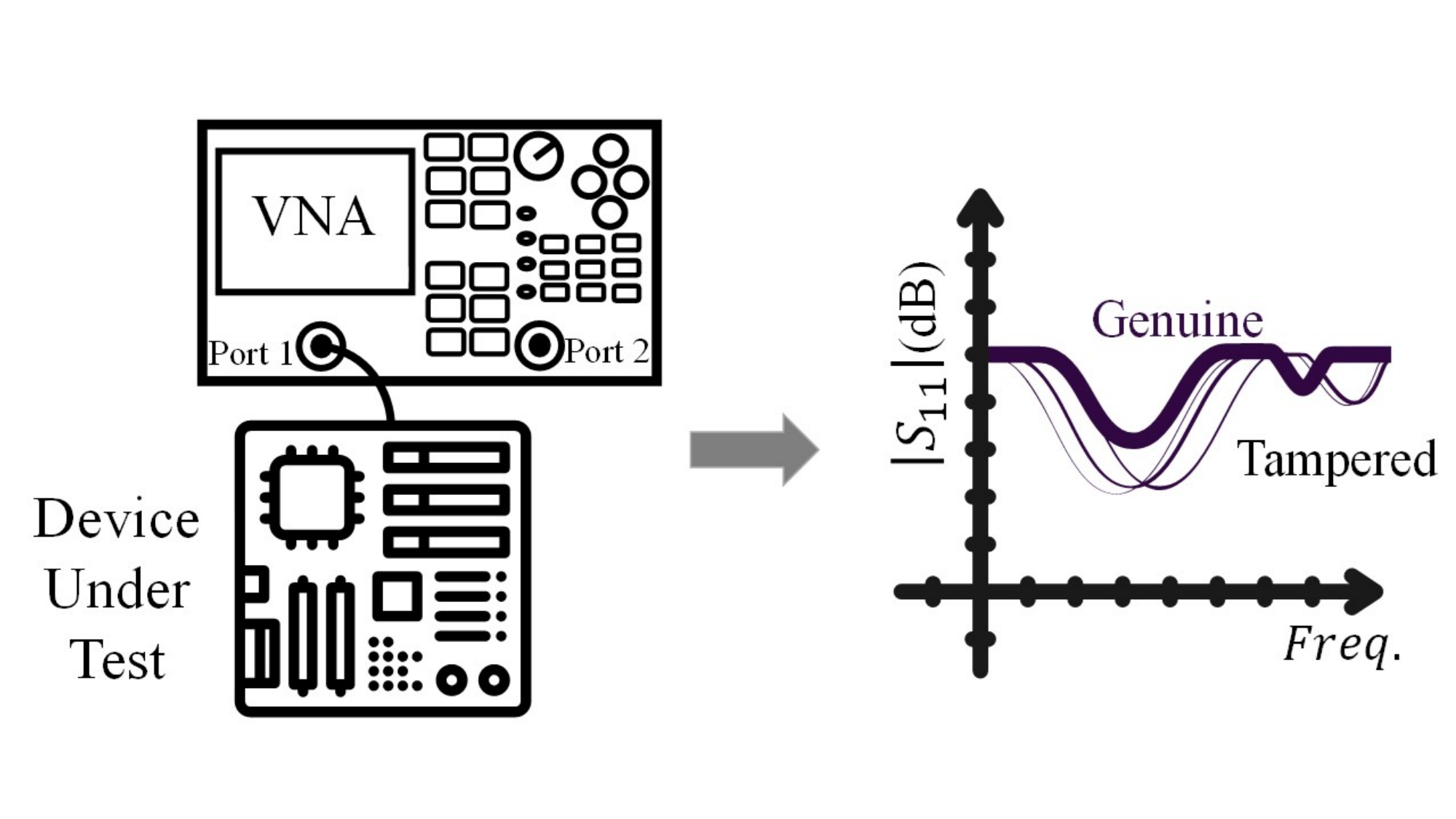}
      \vspace*{-5mm}
     	\caption{Hardware signature extraction using S-parameters.}
     	\vspace{-10pt}
      \label{scattering parameter}
        \end{figure}

Scattering (S-) parameters characterize linear electrical networks when subjected to radio frequency signals.
These parameters are expressed as transmission and reflection coefficients, representing the ratios of transmitted and reflected power to the input power across various frequencies.
PCB's PDN can be modeled as a single-port network.
S-parameters can be obtained through simulations or measured directly in experimental setups using a Vector Network Analyzer (VNA), which accurately determines the network's response over a range of frequencies.
The VNA injects sine waves into the PCB at each frequency sample, capturing the reflective response from the PDN, see Fig.~\ref{scattering parameter}.
The relationship between the impedance of the device under test (DUT),  $Z_{DUT}$, and the reflection coefficient, $S_{11}$, is expressed by $Z_{DUT} = Z_0(1 + S_{11})/(1 - S_{11})$, where $Z_{0}$ represents the characteristic impedance of the cables connecting to the VNA.

\section{Tamper Detection}
\subsection{Threat Model}\label{sec:threat_model}   
\noindent\emph{Attacker}: In our tampering threat model, we assume that attackers can physically tamper with the PCB at any stage of its life cycle, including fabrication, integration, distribution, and repair. 
The adversary has the capability to modify the PCB by adding, removing, replacing, or altering passive components\footnote{Tampering with active components is beyond the scope of this work.}.
Such tampering may lead to counterfeits, clones, malicious functions, or embedded backdoors.
Tamper detection using this method should ideally be conducted before deployment in the field.
However, it is also feasible to detect tampering during the PCB's lifecycle, particularly after distribution. 
To achieve this, the PCB must be disconnected from the system to measure its $|S_{11}|$ signature and verify whether it still matches the genuine signature.
Nevertheless, aging effects can cause gradual changes in the signature over time, making it necessary to study and simulate aging to continuously update the genuine dataset.

\noindent\emph{verifier}: 
Tamper detection is performed by the verifier in two distinct phases: the training phase and the testing phase.
In the training phase, it is assumed that the verifier has full access to the trusted PCB's design files, including the netlist, layout, and bill of materials (BOM).
These files are utilized by the verifier to construct a detailed model of the PCB.
This model is designed to account for potential tampering scenarios by simulating the PCB's behavior under various modifications.
The simulation process can be automated to generate different tampering scenarios and extract the corresponding signatures efficiently.
This simulated data is then used to train a machine learning model, e.g., an RF classifier, which learns to differentiate between the signatures of genuine and tampered PCBs.

In the testing phase, the verifier conducts the verification process on the suspicious PCB.
The verifier has the PCB under test and has access to a VNA to measure the $|S_{11}|$ parameter.
In this treat model, the verifier can use simulated signatures for training. 
To implement this, the pattern of a measured golden signature should be compared with a simulated one using similarity metrics, e.g., Dynamic Time Warping (DTW)~\cite{saadat2024parasitic}. 
Afterward, the obtained metric should be used to align all simulated signatures accordingly, ensuring consistency with the measured data before they are fed into the training model.


\subsection{Tamper Detection and Explanation}\label{sec:tamper_detection_model} 
Our methodology focuses on leveraging $|S_{11}|$ signatures, which provide detailed insights into the electrical behavior of PCBs, including impedance, signal reflections, and potential defects. 
This information is crucial for understanding the circuit layout and identifying anomalies that may indicate tampering.
While simple circuits can be assessed with analytical equations, the complex electromagnetic interactions within modern multi-layered PCBs necessitate the use of more advanced analysis techniques.


The dataset containing the $|S_{11}|$ signatures is a complex multivariate time series dataset. 
This is due to the fact that the impedance, and consequently, $|S_{11}|$ parameter, are spread over consecutive frequency samples; hence, the sample points in each signature are not independent of one another.  
Statistical methods often rely on assumptions of linearity and independence and are, thus, not suitable~\cite{james2013introduction}.  
Each signature exhibits multi-dimensional interactions across data points collected one after another with a given frequency step, which can reflect the characteristics of a multivariate time series.  
Moreover, with high-dimensional time series data, traditional statistical approaches may fail to effectively capture the underlying patterns, resulting in weak predictive performance~\cite{hastie2009elements}.

Machine Learning (ML) algorithms, on the other hand, are well-suited for handling complex datasets with large numbers of data points and multiple interdependent variables~\cite{murphy2012machine}. For instance, models like RFs or Support Vector Machines (SVM) are specifically designed to uncover non-linear relationships and complex interactions without requiring strict parametric assumptions~\cite{breiman2001random}. 
For our multi-class classification problem, we chose to train an RF model. 
While Neural Networks and other ensemble methods can perform well, especially in more complex scenarios, RF 
provides some distinct advantages, making it particularly well-suited for multi-class classification:  (1) its ability to handle high-dimensional data and large feature spaces without the risk of overfitting;  
(2) unlike other models that may require extensive feature engineering or dimensionality reduction techniques, RF inherently manages feature interactions and importance through its ensemble structure~\cite{breiman2001random};  
(3) while CNNs can also capture feature interactions, especially in data with two-dimensional spatial relationships like images, they are less suited for datasets without such structure; 
(4) its non-parametric nature makes it resilient to outliers and noisy data, which are often present in real-world datasets~\cite{liaw2002classification}. 
Moreover, compared to algorithms like SVM~\cite{cortes1995support} or Neural Networks, RF demands much less effort for hyperparameter tuning and, thus, less computational power~\cite{probst2019hyperparameters}.  

More importantly, RFs provide interpretability, which is critical for applications where model transparency is essential. The model offers straightforward methods to be included in XAI, making it easier to understand which features most influence the predictions~\cite{tuv2009feature}. 
This contrasts with models like deep neural networks, where interpretability is often a challenge due to the black-box nature of their architecture. The ability to gain insights into feature contributions helps us to elaborate more on the primary purpose of this research \cite{breiman2001random}.


\subsection{Explainable AI} \label{Explainable AI} 
SHAP (SHapley Additive exPlanations) is a cooperative game theoretic approach that explains the output of ML models~\cite{molnar2022interpretable}. 
SHAP provides a consistent and interpretable approach to understanding the contribution of individual features in ML models. 
While 
traditional XAI methods, such as LIME\cite{ribeiro2016trust} or feature importance scores, often neglect the interactions between features, SHAP accounts for all feature combinations, providing a more comprehensive explanation \cite{molnar2022interpretable,lundberg2017unified}. 
SHAP fairly attributes the contribution of each feature so that if a feature contributes more to a prediction, it will always receive a higher importance value compared to features contributing less \cite{lundberg2017unified}. 
The method's foundation lies in the additive feature attribution model, where the prediction is represented as a sum of contributions from each feature. The SHAP value $\phi_i$ for feature \( i \) is defined as:

\[
\phi_i = \sum_{S \subseteq N \setminus \{i\}} \frac{|S|! \cdot (|N| - |S| - 1)!}{|N|!} \left[ f(S \cup \{i\}) - f(S) \right],
\]

\noindent where \( S \) represents a subset of features, \( N \) is the set of all features, and \( f \) is the prediction function. The equation ensures that each feature's contribution is fairly distributed based on its marginal contribution across all possible subsets \cite{lundberg2017unified}.

\section{Dataset Generation}\label{sec:dataset}
\subsection{Simulation Setup}
\begin{figure}[t!]
     	\centering \noindent
     	\includegraphics[width=9cm]{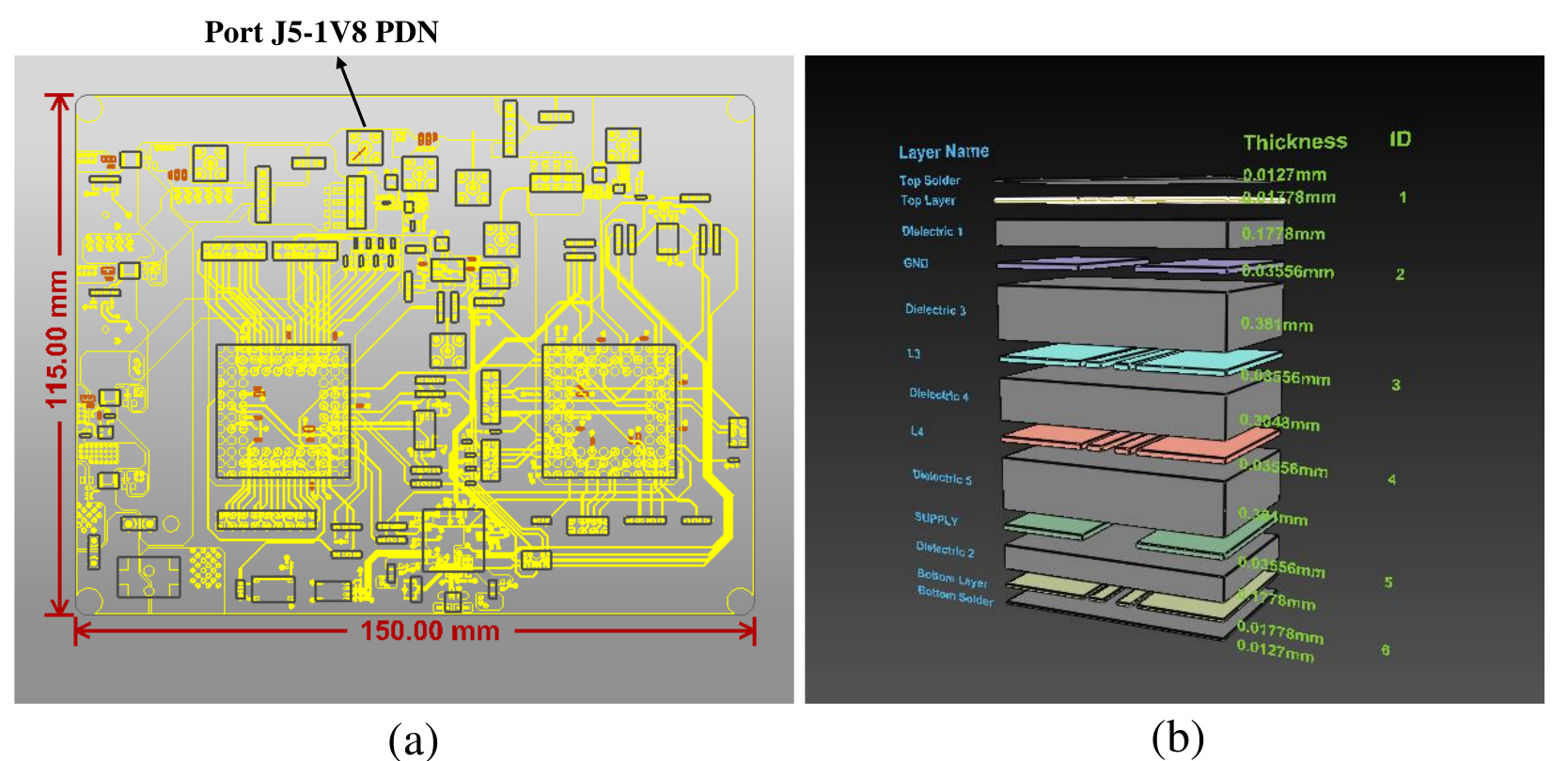}
      \vspace*{-5mm}
     	\caption{(a) Simulated design file of the PCB under test in ANSYS SIwave depicting the circuitry and port J5 of the PDN under test (1V8). (b) stacked-up layers.} 
     \vspace{-10pt}	
      \label{BoardUnderTest}
        \end{figure}

We used an in-house designed PCB which features three separate and isolated PDNs labeled 1V8, 3V3, and 5V, along with six substrate layers.
The PCB, constructed from an FR4 epoxy substrate, comprises 242 components, including capacitors, resistors, ICs, LEDs, SMA ports, headers, traces, and vias.
This study focuses on the 1V8 PDN, with the J5 port providing direct access to the PDN under test, see Fig.~\ref{BoardUnderTest}.

We used Ansys SIwave 2023 R2, a 2.5D electromagnetic (EM) simulation tool that combines the finite element method (FEM) and the method of moments (MOM)~\cite{ansys_siwave_2023}. 
This tool employs a hybrid solver with a 2D triangular mesh, effectively managing intricate PCB layouts.
To achieve accurate results, we dynamically link ANSYS SIwave with HFSS.

\subsection{Data Collection}\label{sec:data_collection}
The first step aimed to detect whether a PCB is genuine or tampered with is to generate datasets for both categories.
One challenge in distinguishing tampered samples from genuine ones is the impact of manufacturing process variations. 
These variations, which refer to deviations in the fabrication process of semiconductor devices, can lead to differences in the physical and electrical properties of the components~\cite{fledell2010pcb}.
These variations can be modeled as changes in parasitic impedance in PCB components, as variations in the dimensions, materials, and placements of metal traces affect the parasitic capacitance, inductance, and resistance. 
Note that real-world PCB components contain parasitics, causing them to behave differently compared to their ideal models.
For instance, in addition to exhibiting capacitive behavior, capacitors also manifest resistive and inductive characteristics, commonly referred to as Equivalent Series Resistance (ESR) and Equivalent Series Inductance (ESL), respectively.
Resistors, in addition to their resistive behavior, also exhibit inductive and capacitive characteristics, known as ESL and Parasitic Capacitance (CP), respectively.
The PDN under test consists of 36 components, including four different types of capacitors with part numbers C0603-CAP-ASM, C0402-CAP-ASM-1, C0805-CAP-ASM, and C0402-CAP-ASM-2, and one resistor with part number 0402-RES-ASM. 
For the genuine dataset, we vary the parasitic values by 10\% around their actual values and 10\% of the component values to model manufacturing process variations.
The $|S_{11}|$ signature from each simulation is exported to a CSV file, resulting in a genuine dataset containing 7500 rows of data across 5000 frequency points from 1 MHz to 1 GHz. This dataset is labeled with a class ID of 0.

To create the tampered dataset, we altered the values of the capacitors and resistors to 10, 100, and 1000 times their original values.
Additionally, we simultaneously varied the parasitic impedances by 20\% around their actual values. 
This process resulted in a CSV file containing 7500 $|S_{11}|$ traces corresponding to tampered PCB samples labeled with a class ID of 1.
In multi-class tampering classification, each tampering attack corresponding to a deviation in the parasitic impedances of each component part number is labeled with a class ID ranging from 1 to 10, as illustrated in Table~\ref{tab:multi_classes}.
In this classification scheme, class ID 0, along with all the tampered classes, are included in the dataset.
By detecting each class, we can determine what has occurred on the PCB and identify which component has been compromised.

\begin{table}[t]
\centering
\scriptsize
\caption{Dataset Classes for Multi-Class Tampering Classification Tasks (G: Genuine, T: Tampered)}
\label{tab:multi_classes}
\resizebox{\columnwidth}{!}{%
\begin{tabular}{|c|c|c|c|c|}
\hline
\textbf{Class ID} & \textbf{Class Name} & \textbf{Value} & \textbf{\# Traces} & \textbf{Description} \\
\hline
0 & C0603-CAP-ASM-G  & 10 uF   & 1500 & $\pm10\%$ C, $\pm10\% \, \text{ESL and ESR}$\\
0 & C0402-CAP-ASM-1-G & 0.1 uF & 1500 & $\pm10\%$ C, $\pm10\% \, \text{ESL and ESR}$ \\
0 & C0805-CAP-ASM-G & 10 uF & 1500 & $\pm10\%$ C, $\pm10\% \, \text{ESL and ESR}$ \\
0 & C0402-CAP-ASM-2-G & 2.2 uF & 1500 & $\pm10\%$ C, $\pm10\% \, \text{ESL and ESR}$ \\
0 & 0402-RES-ASM-G & 470 Ohm & 1500 & $\pm10\%$ C, $\pm10\% \, \text{ESL and CP}$ \\
1 & C0402-CAP-ASM-1-T-ESL & 0.1 uF & 750 &  [10,100,1000]*C, $\pm20\% \, \text{ESL}$\\
2 & C0402-CAP-ASM-1-T-ESR & 0.1 uF & 750 &  [10,100,1000]*C, $\pm20\% \, \text{ESR}$\\
3 & C0402-CAP-ASM-2-T-ESL & 2.2 uF & 750 &  [10,100,1000]*C, $\pm20\% \, \text{ESL}$\\
4 & C0402-CAP-ASM-2-T-ESR & 2.2 uF & 750 &  [10,100,1000]*C, $\pm20\% \, \text{ESR}$\\
5 & C0603-CAP-ASM-T-ESL & 10 uF & 750 & [10,100,1000]*C, $\pm20\% \, \text{ESL}$\\
6 & C0603-CAP-ASM-T-ESR & 10 uF & 750 & [10,100,1000]*C, $\pm20\% \, \text{ESR}$\\
7 & C0805-CAP-ASM-T-ESL & 10 uF & 750 & [10,100,1000]*C,$\pm20\% \, \text{ESL}$\\
8 & C0805-CAP-ASM-T-ESR & 10 uF & 750 & [10,100,1000]*C, $\pm20\% \, \text{ESR}$\\
9 & 0402-RES-ASM-T-CP & 470 Ohm & 750 & [10,100,1000]*R, $\pm20\% \, \text{CP}$ \\
10 & 0402-RES-ASM-T-ESL & 470 Ohm & 750 & [10,100,1000]*R, $\pm20\% \, \text{ESL}$ \\
\hline
\end{tabular}%
}
\end{table}

\subsection{Effect of PDN Components}
During the board simulation and extraction of the $|S_{11}|$ parameter, it was found that the fundamental resonance frequency, identified  by the lowest magnitude, occurs at 470 MHz and is affected by changes in all circuit components.
According to the literature, the PCB impedance exhibits the most significant variation at the resonance frequency because the impedance of the PCB is substantially higher at this frequency~\cite{han2018cipa}.
The impact of each component class on the dataset depends on the component's value, quantity in the PDN, and physical dimensions.
These factors determine the extent of the impact and its location within the frequency domain of the $|S_{11}|$ signature.
Varying the capacitance and ESL values of the capacitors shifts both the fundamental and local resonance frequencies.
In contrast, altering the resistors value and ESR primarily affects the amplitude of the $|S_{11}|$ signature, reflecting its resistive nature.
Notably, during feature extraction, the most significant features or frequency points align closely with those predicted by the physical analysis of each component class.

The majority of components under test in the PDN are C0402-CAP-ASM-1 capacitors, each with a capacitance value of 0.1 uF. 
Although these are the smallest components, their large number makes their effects noticeable across the entire frequency spectrum especially at lower frequencies. 
In the PDN under test, a single C0402-CAP-ASM-2 capacitor with a capacitance of 2.2 uF results in only a minor shift in the fundamental resonance frequency. 
Despite this subtle shift, our model demonstrates a high level of accuracy in detecting these changes (see Section~\ref{results}). 

\begin{figure}[t!]
     	\centering \noindent
     	\includegraphics[width=9cm]{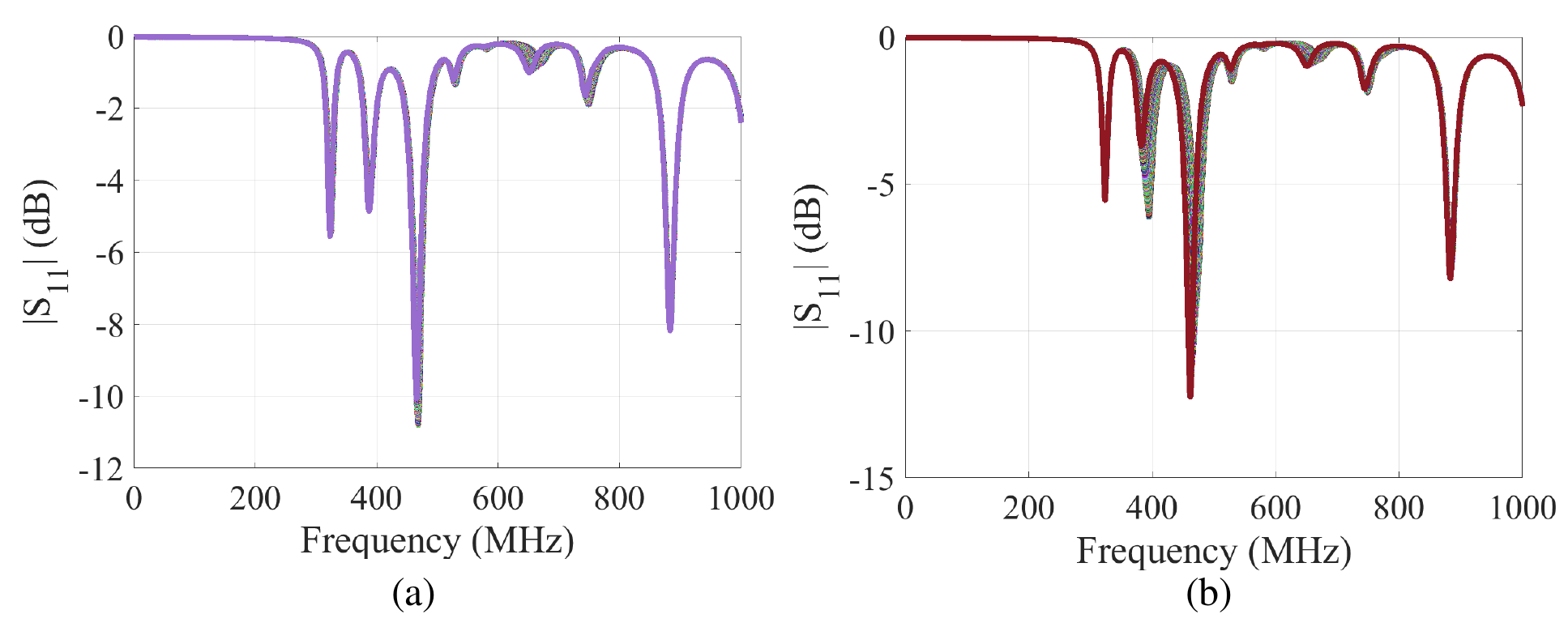}
      \vspace*{-5mm}
     	\caption{(a) S-parameter traces corresponding to C0603-CAP-ASM-T-ESL class. (b) S-parameter traces corresponding to C0805-CAP-ASM-T-ESL class.}
     	\label{dataset_presentation}
        \end{figure}

An interesting observation is that the C0805-CAP-ASM capacitors, which are the largest physically, primarily influence local resonance frequencies below the fundamental resonance frequency when their values and ESL are changed.
Specifically, this effect is observed around 395 MHz, which aligns with the most important features reported by the feature extraction process.
In contrast, the C0603-CAP-ASM, despite having the same capacitance values, influences frequencies higher than the fundamental resonance frequency,  due to its smaller physical dimensions compared to the C0805-CAP-ASM capacitors as shown in Fig.~\ref{dataset_presentation}.
In this scenario, the local resonance frequencies that are shifted are approximately 662 MHz and 712 MHz.
The most significant feature numbers for this component class correspond closely with the affected local resonance frequencies. 
This difference arises because smaller capacitors significantly impact impedance at higher frequencies due to their smaller physical dimensions, causing resonance at those higher frequencies~\cite{mosavirik2023impedanceverif}.
The effects of classes 9 and 10 have been observed across all resonance frequencies, as the parasitic capacitance and parasitic inductance added to the circuit both cause shifts in resonances.
It is observed that in some classes, multiple features corresponding to very low frequencies are particularly important.
This can be attributed to the fact that, in scattering parameter analysis, the effect of resistance is most pronounced at lower frequencies, which aligns with changes in the parasitic resistance within these classes. 
At lower frequencies, resistive elements dominate the impedance characteristics, directly affecting the S-parameters.
As the frequency increases, reactive components like inductors and capacitors begin to have a greater influence, diminishing the impact of resistance.
Therefore, variations in scattering parameters at lower frequencies are primarily driven by resistive elements, making these frequencies crucial for understanding changes in parasitic resistance.

\section{Results}\label{results}



\subsection{Experimental Setup}\label{Experimental Setup}
The experiments were conducted on a high-performance computing system running Ubuntu 22.04.3 LTS with a 6.5.0-44-generic Linux kernel. The system was equipped with dual Intel(R) Xeon(R) Silver 4216 CPUs, each with 16 cores and 32 threads, providing a total of 64 logical processors. The system had 256 GB of RAM. For GPU-accelerated computations, an NVIDIA RTX A4000 GPU with 16 GB of dedicated memory was utilized. 
Our dataset of $|S_{11}|$ signatures is characterized by 5,000 samples per trace collected over a frequency range as explained in Section~\ref{sec:data_collection}. 

\subsection{Feature Engineering}\label{Feature Engineering} 
As described in Section~\ref{sec:tamper_detection_model}, our multi-class classification problem is tackled by training an RF model at the core of XAI. 
Before training the model, it is tempting to apply pre-processing techniques to reduce the number of features needed to enhance training. 
It might be thought that the commonly-applied Principal Component Analysis (PCA) would be beneficial. 
PCA effectively captured the variance in the data with a reduced number of components while maintaining good predictive performance. 
However, despite these benefits, using PCA completely confused SHAP. 
This limitation arises because each PCA component is a linear combination of multiple original features, making it impossible to isolate and interpret the contribution of individual features. Consequently, this limited our ability to provide clear explanations of feature importance, essential for understanding the underlying factors driving model predictions.
Another orthogonal idea is to standardize the features to center the data around the mean and scale it to unit variance to reduce the impact of the extreme values. 
This ensures that the transformed data is less sensitive to outliers, thereby enhancing the robustness of our ML model. 
We applied this transformation before training the model. 

\begin{table}[t]
\scriptsize
\centering
\caption{Accuracy of each fold involved in $k=5$-fold cross-validation and the characteristics of the RF trained on that fold.}
\label{tab:Accuracy}
\begin{tabular}{|c|c|c|c|}
\hline
Fold       & Accuracy         & Minimum depth & Maximum depth \\ \hline
1          & 96.33\%          & 33            & 49            \\ \hline
2 & 96.83\% & 34   & 54   \\ \hline
3          & 96.53\%          & 34            & 54            \\ \hline
4          & 96.70\%          & 34            & 50            \\ \hline
5          & 96.70\%          & 35            & 53            \\ \hline
\end{tabular}
\end{table}
\subsection{Model Training}\label{Model Training}
After pre-processing, we trained our RF model by using 5-fold cross-validation. 
The number of folds ($k=5$) is selected by conducting the commonly applied sensitivity analysis for $2\leq k \leq 10$, where the mean accuracy obtained for $k$ is compared to mean classification accuracy from leave-one-out cross-validation (LOOCV) on the same dataset. 
Based on this, we observed that for $k\geq 5$, the sensitivity converges to its minimum; hence, we chose $k=5$. 
The accuracy for each of the 5 folds and the characteristics of the RF trained on each fold are listed in Table~\ref{tab:Accuracy}. 
It can be seen that the variation in accuracy and characteristics of the RF models is statistically insignificant, i.e., the trained model is robust. 
In addition to accuracy, we also checked the confusion matrices and observed that classes are being predicted equally well and no classes are neglected by the model. 
Note that unknown modifications that were not labeled during training can be detected as non-genuine; however, the model cannot determine the exact nature of the changes.


\begin{figure*}[htbp]
    \centering
    \resizebox{\textwidth}{!}{
        \begin{minipage}{\textwidth}
            \centering
            \includegraphics[width=0.48\textwidth]{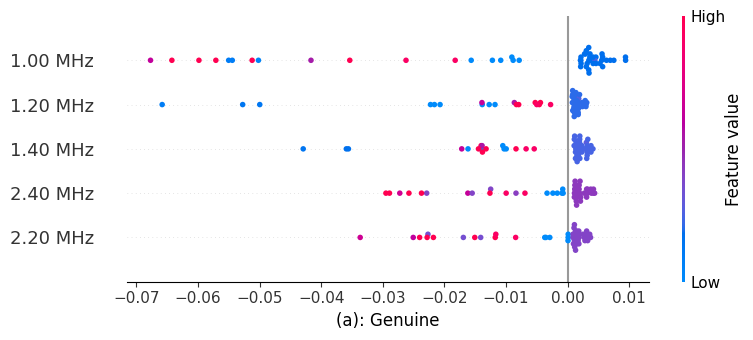} \hfill
            \includegraphics[width=0.48\textwidth]{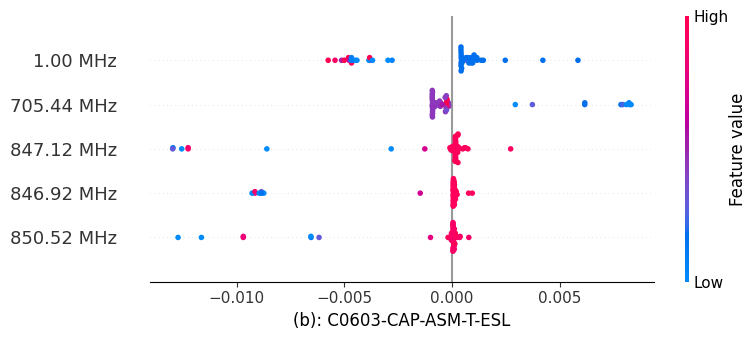} \\
            \vspace{1em}
            \includegraphics[width=0.48\textwidth]{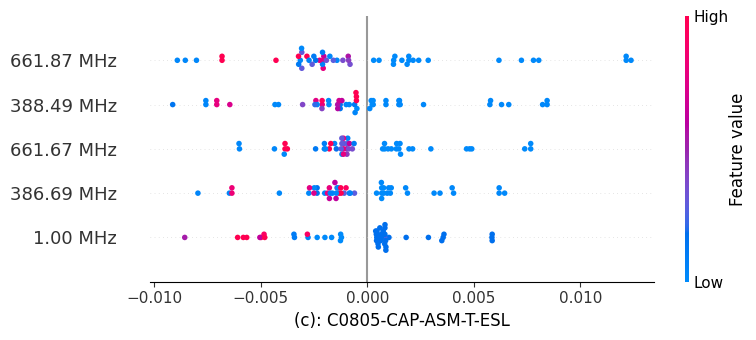} \hfill
            \includegraphics[width=0.48\textwidth]{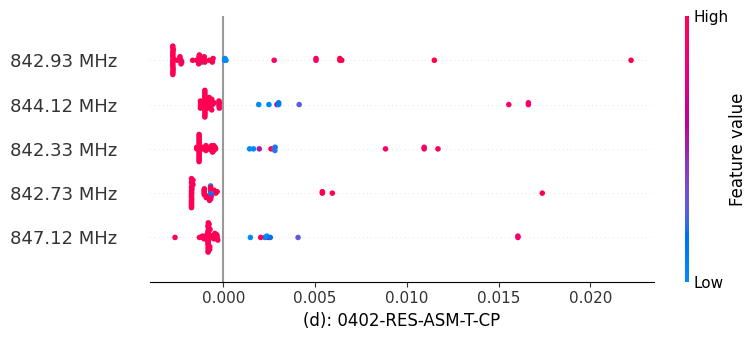} \\
            \vspace{1em}

        \end{minipage}    }
    \caption{SHAP values: Each dot represents a sample, with its position on the x-axis indicating the impact of a particular signature feature collected at a frequency (e.g., 4.20 MHz, 2.40 MHz) on the prediction. The sign of the values on the x-axis indicates whether a sample influences the prediction positively or negatively. The color gradient from blue to red shows the feature value, where blue is for lower values and red for higher values.}
    \label{SHAP results}
\end{figure*}

\subsection{Results for XAI through SHAP}
To show the contribution of each feature in the classification task, we calculated the SHAP values of 50 randomly chosen test samples, i.e., the default value for TreeSHAP implemented in the SHAP Python package~\cite{shap}. We also increased the number of samples per class to 100 and 200; however, the results have been the same. Hence to accelerate the analysis, 50 samples per class are employed. This hyperparameter may affect the output of the explainer, although in our case, we did not observe any significant change when tuning that. 
In figure \ref{SHAP results}, we present the SHAP values of the four classes as an example (see the summary of results for all classes in Table~\ref{tab:shap}). This figure illustrates the contribution of the different features to the prediction of the classes. 
The distribution reveals that certain frequencies, such as 1.00 MHz and 1.2 MHz for the class ``Genuine'', have more impacts on the model's decision. 
Clusters of SHAP values show regions where feature contributions push predictions towards or away from the ``Genuine'' outcome.  
In the ``Genuine'' class, the 1.00 MHz, 1.2 MHz, and 2.40 MHz frequencies show more contributions, with values widely dispersed from zero, indicating both positive and negative impacts on the prediction. Similarly, for the C0603-CAP-ASM-T-ESL class, features such as 705.44 MHz, 847.12 MHz, and 846.92 MHz have significant spreads, suggesting their strong roles in affecting the model's output.
The lower plots for C0805-CAP-ASM-T-ESL and C0402-RES-ASM-T-CP highlight different sets of frequency features, such as 388.49 MHz and 842.93 MHz, which also vary in their importance based on their position relative to the zero impact line. Notably, some frequencies show consistent impacts across classes, while others affect only some classes.

\begin{table}[t]
\scriptsize
\centering
\caption{SHAP values of the most important features}\label{tab:shap}
\begin{tabular}{|c|c|c|c|}
\hline
Class ID & Most important feature & Mean  & Variance \\ \hline
0        & 1 MHz                  & -8.18e-5             & 7.87e-7                  \\ \hline
1        & 1MHz                   & 1.04e-5               & 2.91e-7                 \\ \hline
2        & 1MHz                   & 6.29e-6              & 1.81e-7                  \\ \hline
3        & 1MHz                   & 9.89e-6             & 2.16e-7                 \\ \hline
4        & 1MHz                   & 1.01e-5              & 1.58e-7                 \\ \hline
5        & 1.00MHz                & -8.27e-7             & 1.14e-7                 \\ \hline
6        & 661.87MHz              & 4.75e-6            & 1.12e-7                 \\ \hline
7        & 661.87MHz              & 1.59e-5               & 1.24e-7                 \\ \hline
8        & 473.62MHz              & 1.59e-6             & 7.54e-8                 \\ \hline
9        & 842.93MHz              & 9.63e-6             & 1.42e-7                 \\ \hline
10       & 845.52MHz              & 1.4e-5               & 3.02e-7                 \\ \hline
\end{tabular}
\end{table}

\section{Conclusion}
We proposed a novel XAI framework to enable tamper forensic analysis of PCBs.
We formulated tamper detection as a supervised ML task. To enhance the robustness of our approach, we simulate various tampering scenarios to generate two-dimensional impedance signatures, which are then used to train an RF model. 
This enables accurate classification of different tamper types and precise localization of tamper events. 
While we validated our claims by considering replacing counterfeit components, the method is applicable to component insertion or removal as well. 
Our method not only offers highly accurate detection but also explains which characteristics of the signatures contribute to classifying the tamper types, and thus, it reduces the need for time-consuming manual inspections. 

\noindent\textbf{Future Work.} Applying explainability to chip-level tamper detection is an interesting future research direction. While it has been shown that chip-level tamper detection using PDN impedance analysis is feasible~\cite{mosavirik2023silicon}, further investigation is still required to explore whether the proposed scheme in this work is extendable to chip-level tamper events. Finally, using the on-chip VNAs,~\cite{mosavirik2023impedanceverif} to perform explainable tamper detection is another major future research direction, as it removes the need for an external VNA and human intervention. 
 \section*{Acknowledgment}
This effort was sponsored in part by NSF Grants CNS-2338069, ECCS-2138420, and DGE-2021871 and in part by EPRI.

\bibliographystyle{ieeetr}
\bibliography{references}

\end{document}